\documentclass{aa}  

\usepackage[titletoc,title]{appendix}

\usepackage{graphicx}
\usepackage{txfonts}
\usepackage{caption}
\usepackage{amsmath} 
\usepackage{subfigure}

\begin{document} 
   \title{Do Some AGN Lack X-ray Emission?}

   \author{ C. Simmonds\inst{1}
          \and
          F. E. Bauer\inst{1,2,3} 
          \and
          T. X. Thuan \inst{4}
          \and 
          Y. I. Izotov \inst{5}
          \and 
          D. Stern \inst{6}
          \and 
          F. A. Harrison\inst{7}
          }
   \institute{Instituto de Astrof{\'{\i}}sica and Centro de Astroingenier{\'{\i}}a, Facultad de F{\'{i}}sica, Pontificia Universidad Cat{\'{o}}lica de Chile, Casilla 306, Santiago 22, Chile\\
              \and 
              Millennium Institute of Astrophysics (MAS), Nuncio Monse{\~{n}}or S{\'{o}}tero Sanz 100, Providencia, Santiago, Chile\\ 
              \and 
              Space Science Institute, 4750 Walnut Street, Suite 205, Boulder, Colorado 80301\\
              \and 
              Astronomy Department, University of Virginia, P.O. Box 400325, Charlottesville, VA 22904-4325\\
              \and 
              Main Astronomical Observatory, Ukrainian National Academy of Sciences, 27 Zabolotnoho Street, Kyiv 03680, Ukraine \\
              \and 
              Jet Propulsion Laboratory, California Institute of Technology, Pasadena, CA 91109, USA\\
         \and
              Cahill Center for Astronomy and Astrophysics, California Institute of Technology, Pasadena, CA 91125, USA
             }

   \date{Received Month XX, 2016; accepted Month XX, 2016}

 
  \abstract
   {Intermediate-Mass Black Holes (IMBHs) are thought to be the seeds of early Supermassive Black Holes (SMBHs). While $\gtrsim$100 IMBH and small SMBH candidates have been identified in recent years, few have been robustly confirmed to date, leaving their number density in considerable doubt. Placing firmer constraints both on the methods used to identify and confirm IMBHs/SMBHs, as well as characterizing the range of host environments that IMBHs/SMBHs likely inhabit is therefore of considerable interest and importance. Additionally, finding significant numbers of IMBHs in metal-poor systems would be particularly intriguing, since such systems may represent local analogs of primordial galaxies, and therefore could provide clues of early accretion processes.}
   {Here we study in detail several candidate Active Galactic Nuclei (AGN) found in metal-poor hosts.}
   {We utilize new X-ray and optical observations to characterize these metal-poor AGN candidates and compare them against known AGN luminosity relations and well-characterized IMBH/SMBH samples.}
   {Despite having clear broad optical emission lines that are long-lived ($\gtrsim$10--13\,yr), these candidate AGN appear to lack associated strong X-ray and hard UV emission, lying at least 1--2 dex off the known AGN correlations. If they are IMBHs/SMBHs, our constraints imply that they either are not actively accreting, their accretion disks are fully obscured along our line-of-sight, or their accretion disks are not producing characteristic high energy emission. Alternatively, if they are not AGN, then their luminous broad emission lines imply production by extreme stellar processes. The latter
would have profound implications on the applicability of broad lines for mass estimates of massive black holes.}
   {}



   \maketitle
%

\section{Introduction}

Active Galactic Nuclei (AGN) are usually found in massive, bulge-dominated galaxies that have already converted most of their gas into stars by the present epoch, and consequently tend to have high metallicities. Observations of AGN have borne this out, showing that AGN hosts usually possess metallicities ranging from solar to supersolar \citep[e.g.,][]{storchi-bergmann,hamann}. This raises the question as to whether low-metallicity AGN exist, and if so, in what types of galaxies? We can glean some insight from the black hole (BH) mass to bulge luminosity relation \citep{kormendy,magorrian,bentz} and the BH mass to bulge velocity dispersion relation \citep[$M_{\rm BH}-\sigma$;][]{haehnelt,gebhardt,ferrarese,tremaine,greene06}, both of which relate BH mass to galaxy growth. These relations have largely been established only in nearby massive galaxies \citep[with $M_{\rm BH}$$\sim$$10^{6}$--$10^{10}$\,$M_{\odot}$; e.g.,][]{McConnell2013}, where a BH's gravitational influence can be resolved and studied, but we expect that they should extend to higher and lower galaxy mass regimes. 

Although there is still much debate about how these relations behave in the low-mass regime \citep[e.g.,][]{barth05,greene10,jiang,graham13,sartori}, these relations naively imply that intermediate-mass BHs (IMBHs), which are thought to be the missing link between stellar-mass and supermassive BHs (SMBHs) and potential seeds of SMBHs in the early universe \citep[e.g.,][]{volonteri}, should occur in low-mass dwarf galaxies. The discovery and characterization of IMBHs are thus of particular interest. 

The most robust way to confirm the presence of a BH is via spatially resolved dynamical estimates. However, this method is difficult to employ in practice due to the small regions of influence around IMBHs. Thus it is often necessary to resort to indirect estimates. One method is to search for the telltale signs of broad-line emission associated with AGN activity due to photoionized gas within the gravitational influence of the BH. \citet{greene04, greene06}, \citet[hereafter I07]{izotov07}, \citet{reines13} and others, for instance, have used the SDSS spectroscopic archives to systematically search for such broad-line tracers, finding 10's to 100's of candidates. A second method is to use diagnostic line ratios that assess the underlying UV spectrum to separate out AGN activity from star formation and shocks \citep[e.g.,][]{groves,reines13,moran}. Finally, a third method is to search for X-ray, mid-IR, and/or compact, low-luminosity radio emission, which can sometimes be interpreted as unambiguous signs of AGN activity \citep[e.g.,][]{greene06,reines14,Hainline2016}; notably, 70\% of the \citet[][hereafter GH04]{greene04} sample have clear X-ray emission, strengthening their identification \citep[][hereafter D09]{desroches}. Despite the discovery of numerous AGN candidates in dwarf galaxies with these methods, few IMBHs with masses below 10$^{5}$ $M_{\odot}$ have been conclusively identified (e.g., NGC 4395, RGG118 and HLX-1) and few AGN overall have been found in metal-poor ($Z$$<$$0.2Z_{\odot}$) systems which might be considered analogs of early AGN hosts \citep{groves}.\\

In this work, we study several metal-poor systems which have been argued to host candidate AGN based on luminous, constant, broad emission lines. We place new constraints on these objects using X-ray observations from the {\it Chandra} Observatory and additional ground-based optical spectroscopy. The outline of the paper is as follows. In $\S$2, we describe the X-ray and optical observations and data reduction procedures. In $\S$3, we present the results and compare these to other known AGN, as well as possible alternative interpretations. In $\S$4, we summarize our work and discuss broader implications.\\

   \begin{table*}
   \begin{center}
   {\small
       \caption[]{General Properties of Sample}
         \label{general}
         \begin{tabular}{ccccccccccc}
           	\hline
           	\noalign{\smallskip}
           	Object & R.A. & Dec. & $z$ & $L_{\rm H\alpha, br}$ & $L_{\rm H\alpha, br}$/$L_{\rm H\alpha, nar}$ & $L_{\rm [O{\sc III}]}$ & $g$ & $M_{g}$ & $12 +\log{\rm O/H}$ & $M_{\rm BH}$ \\
           	\noalign{\smallskip}
           	\hline
           	\noalign{\smallskip}
           	J0045+1339 & 00 45 29.2 & +13 39 09 & 0.29522 & 2.7$\times$10$^{41}$ & 0.42 & 3.0$\times$10$^{42}$ &21.8 & -18.6 & 7.9 & 2.4$\times$10$^{6}$ \\
           	J1025+1402 & 10 25 30.3 & +14 02 07 & 0.10067 & 3.2$\times$10$^{41}$ & 3.38 & 2.7$\times$10$^{41}$ & 20.4 & -17.7  & 7.4 & 5.1$\times$10$^{5}$ \\
           	J1047+0739 & 10 47 55.9 & +07 39 51 & 0.16828 & 1.6$\times$10$^{42}$ & 1.22 & 3.6$\times$10$^{42}$ & 19.9 & -19.2 & 8.0 & 3.1$\times$10$^{6}$ \\
           	J1222+3602 & 12 22 45.7 & +36 02 18 & 0.30112 & 2.8$\times$10$^{41}$ & 0.72 & 2.0$\times$10$^{42}$ & 21.3 & -19.1 & 7.9 & 6.3$\times$10$^{5}$ \\
            J1536+3122 & 15 36 56.5 & +31 22 48 & 0.05619 & 4.6$\times$10$^{40}$ & 0.19 & 1.8$\times$10$^{39}$ & 17.5 & -19.3 & 8.3 & 3.3$\times$10$^{5}$ \\
            J0840+4707 & 08 40 29.9 & +47 07 10 & 0.04219 & 8.5$\times$10$^{40}$ & 0.17 & 8.5$\times$10$^{39}$ & 17.6 & -18.5 & 7.6 & 3.0$\times$10$^{5}$ \\
            J1404+5423 & 14 04 28.6 & +54 23 53 & 0.00117 & 1.9$\times$10$^{38}$ & 0.15 & 7.4$\times$10$^{36}$ & 16.7 & -12.4 & 7.9 & 4.2$\times$10$^{4}$? \\
           	\noalign{\smallskip}
           	\hline
       	\end{tabular}
       \tablefoot{
       {\it Col 1}: Shortened SDSS object name, from I07.
       {\it Col 2}: Right ascension in epoch J2000.0.
       {\it Col 3}: Declination in epoch J2000.0.
       {\it Col 4}: Redshift.
       {\it Col 5}: Luminosity of the broad H$\alpha$, $L_{\rm H\alpha, br}$, in erg s$^{-1}$.
       {\it Col 6}: Ratio of broad/narrow H$\alpha$ luminosity components, $L_{\rm H\alpha, br}$/$L_{\rm H\alpha, nar}$.
       {\it Col 7}: Luminosity of the [O{\sc iii}] $\lambda$5007, $L_{\rm [O{\sc III}]}$, in erg s$^{-1}$.
       {\it Col 8}: Apparent magnitude, $g$, from SDSS, in ABmag. 
       {\it Col 9}: Absolute magnitude, $M_{g}$, in ABmag.
       {\it Col 10}: Metallicity based on oxygen abundance, $12 + \log{\rm O/H}$.
       {\it Col 11}: Black hole mass, $M_{\rm BH}$, estimated from H$\alpha$ relation of GH05, in $M_{\odot}$. The last object, SDSS J1404+5423, is located $\sim$11\arcmin\ ($\sim$16\,kpc) from the nucleus of M101, which likely rules out a central massive black hole. 
       }
       }
   \end{center}
   \end{table*}

   \begin{table}
   \begin{center}
   \caption[]{Multi-epoch Broad H$\alpha$ Fluxes for AGN Candidates}
         \label{flux}
         \begin{tabular}{cccc}
           	\hline
           	\noalign{\smallskip}
           	Object & $F_{\rm H\alpha, br}$ & Obs. Date & Telescope \\ 
            \noalign{\smallskip}
           	\hline
           	\noalign{\smallskip}
           	J0045+1339 & 16.4$\pm$1.7  & 2000 Jan 12 & SDSS \\
                       & 18.0$\pm$0.7  & 2007 Nov 15 & APO  \\
                       & 20.4$\pm$0.4  & 2010 Feb 07 & APO  \\
                       & 25.2$\pm$0.5 & 2013 Oct 27 & LBT  \\
            J1025+1402 & 165.0$\pm$5.7 & 2004 Mar 11 & SDSS \\
                       & 192.5$\pm$0.8 & 2008 Feb 06 & APO  \\
                       & 227.1$\pm$4.5 & 2009 Nov 20 & APO  \\
                       & 260.3$\pm$3.3 & 2010 Feb 07 & APO  \\
                       & 276.8$\pm$6.1 & 2015 Feb 24 & APO  \\
                       & 185.9$\pm$3.7 & 2016 Feb 14 & Keck \\
            J1047+0739 & 289.2$\pm$9.1 & 2003 Jan 31 & SDSS \\
                       & 224.2$\pm$2.1 & 2008 Feb 06 & APO  \\
                       & 190.5$\pm$3.9 & 2009 Nov 20 & APO  \\
                       & 208.0$\pm$4.3 & 2010 Feb 07 & APO  \\
                       & 233.7$\pm$5.1 & 2015 Feb 24 & APO  \\
                       & 215.7$\pm$4.3 & 2016 Feb 14 & Keck \\
            J1222+3602 & 16.1$\pm$1.8  & 2005 Mar 13 & SDSS \\
                       & 22.4$\pm$1.3  & 2008 Feb 06 & APO  \\
                       & 27.0$\pm$0.6  & 2015 May 18 & LBT \\
            J1536+3122 & 78.7$\pm$4.1  & 2004 Apr 24 & SDSS \\
            J0840+4707 & 244.8$\pm$7.5 & 2001 Mar 13 & SDSS \\
            J1404+5423 & 350.4$\pm$10.6 & 2004 Mar 24 & SDSS \\
\noalign{\smallskip}
           	\hline
       	\end{tabular}
          \tablefoot{
          {\it Col. 1}: Shortened SDSS object name.
          {\it Col. 2}: Flux of H$\alpha$ broad component in units of 10$^{-16}$ erg\,s$^{-1}$ cm$^{-2}$. Only statistical errors are reported.
          {\it Col. 3}: Observation date (UT). 
          {\it Col. 4}: Telescope.
          }
   \end{center}
   \end{table}

\section{Data}\label{sec:data}

\subsection{Sample}

While searching for low-metallicity dwarf galaxies in the Sloan Digital Sky Survey (SDSS, DR5), I07 discovered a subset of 21 emission-line galaxies (ELGs) which have broad H$\alpha$ emission suggesting potential AGN activity. From these, \citet[][hereafter I08]{izotov08} singled out four of the most extreme metal-poor ($Z$$=$0.05--0.16\,$Z_{\odot}$) sources as having the strongest evidence for hosting candidate AGN based on their very unusual spectral properties. 

We targeted with dedicated \textsl{Chandra} observations the four most extreme ELGs reported by I08 (J1025$+$1402, J1047$+$0739, J0045$+$1339, and J1222$+$3602). For completeness, we also report the results of archival \textsl{Chandra} observations for three weaker objects (J1536$+$3122, J0840$+$4707, and J1404$+$5423) in the parent I07 sample which also have broad H$\alpha$ emission lines. Table \ref{general} presents the names and coordinates of all seven metal-poor ELGs. 

The four I08 ELGs all have strong, broad permitted H$\alpha$ $\lambda$6563 lines (Full Width at Half Maximum (FWHM)$\approx$1600--1900\,km\,s$^{-1}$, Full Width at Zero Intensity (FWZI)$\approx$2200--3500\,km\,s$^{-1}$) with luminosities $L_{\rm H\alpha, br}$$=$3$\times$10$^{41}$--2$\times$10$^{42}$\,erg\,s$^{-1}$ which persist with little variability over periods of at least $\sim$3--7 years (I08). Such lines are far too broad, luminous, and constant to be attributed to stellar processes such as Wolf-Rayet stars, Ofp or luminous blue variable star winds, single or multiple supernova (SN) remnants, giant SN bubbles, or even shocks propagating in dense circumstellar envelopes around Type IIn SNe. Standard stellar mechanisms can typically only account for $L_{\rm H\alpha,br}$$\sim$$10^{36}$--$10^{40}$ erg\,s$^{-1}$ \citep[e.g.,][]{kennicutt1989,gunawardhana}, while the I08 objects are 30 to 200 times more luminous. Type IIn SNe can potentially generate $L_{\rm H\alpha,br}$$\sim$$10^{38}$--$10^{42}$ erg s$^{-1}$, but only for a few years in the most extreme cases \citep[e.g., SN\,2005ip and SN\,2006jd;][]{stritzinger} and generally have notable variability (see $\S$\ref{sec:discussion}). This leaves only an AGN origin, although extremely luminous and long-lived SNe cannot be completely ruled out. For the three weaker I07 objects (FWHM$\approx$1300--1500\,km\,s$^{-1}$, FWZI$\approx$3300--6700\,km\,s$^{-1}$ with luminosities $L_{\rm H\alpha, br}$$=$2$\times$10$^{38}$--9$\times$10$^{40}$\,erg\,s$^{-1}$), the situation is more ambiguous, since all mechanisms, stellar and non-stellar, remain viable.

\begin{figure*}
   \centering
   \includegraphics[width=15cm]{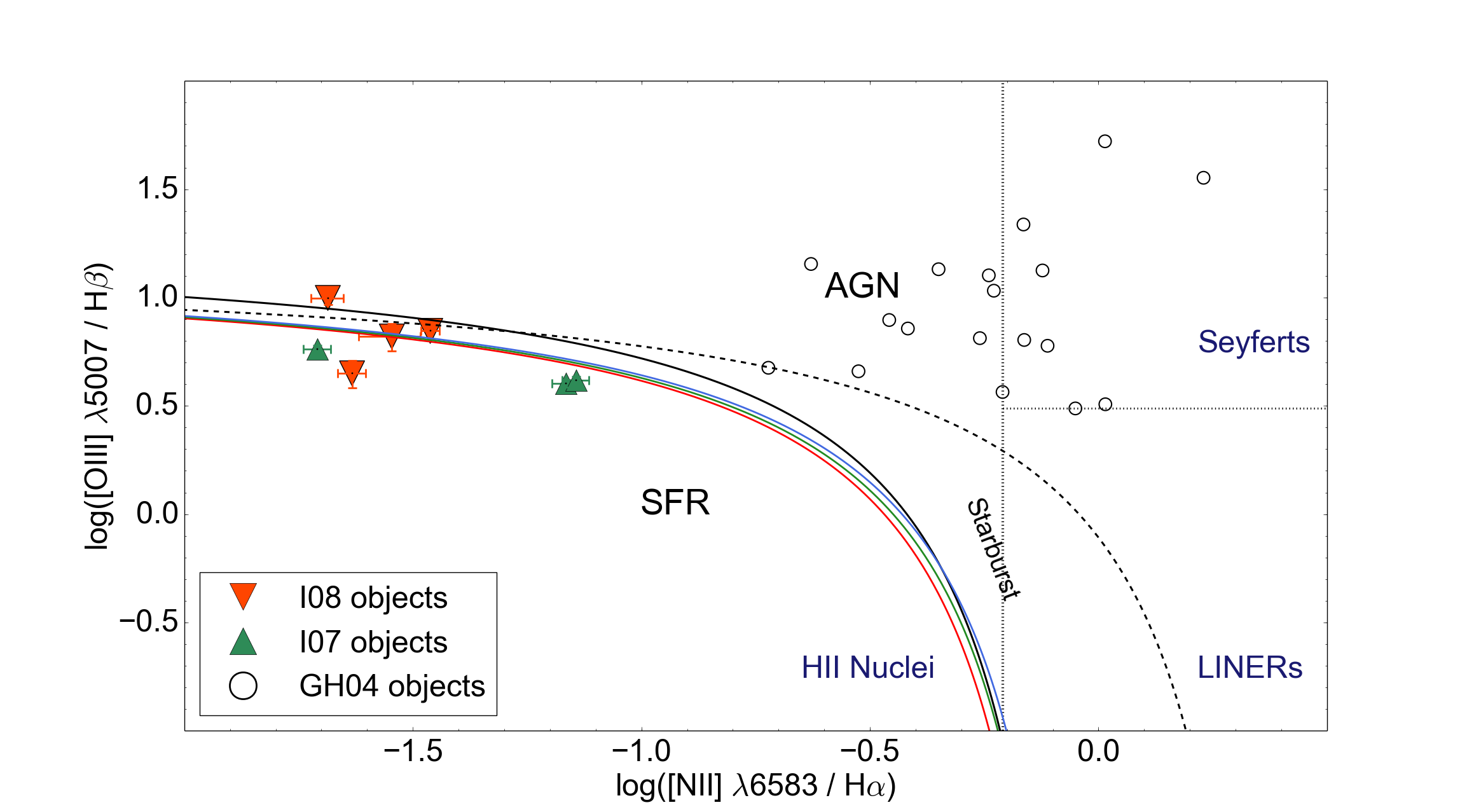}
      \caption[BPT Diagram]{[O{\sc iii}]/H$\beta$ against [N{\sc ii}]/H$\alpha$ narrow line emission ratios. Vertical and horizontal dotted lines mark the boundaries of the three main classes of emission line nuclei from \citet{ho}. The dashed curve shows the demarcation between starburst galaxies and AGNs defined by \citet{kewley01}. The solid black curve shows the demarcation between star-forming H{\sc ii} regions and starburst galaxies from \citet{kauffmann}, while the solid red, blue and green curves denote the AGN-H{\sc ii} separation from \citet{kewley13}, corrected to redshifts of $z$$=$0.1, 0.2 and 0.3, respectively. Filled red and green triangles represent the I08 and I07 objects, respectively. Open circles represent the GH04 objects. The I08 and I07 galaxies lie on the left side near the boundary separating AGN from star-forming regions. Their location is consistent with the stellar origin, but non-thermal radiation is not excluded. It is important to note that metal-poor AGN will have weaker [N{\sc ii}] and [O{\sc iii}] emission, such that they migrate from the right side of the diagram to the left \citep[e.g.,][]{groves}, making the standard diagnostic more ambiguous. Thus, the origin of the ionising process cannot be determined by this diagram alone.
     }
     \label{BPT}
   \end{figure*}

All of the objects closely follow the AGN relations which relate $L_{\rm 5100\AA}$ to
both $L_{\rm H\alpha, br}$ and $L_{\rm [OIII]}$ \citep[e.g.,][hereafter GH05]{greene05}. However, this only implies that the observed UV/optical continuum is sufficient to power the H$\alpha$ and [O{\sc iii}] emission and does not necessarily indicate what the emission mechanism is (e.g., AGN accretion). If we look at the AGN/star formation emission-line diagnostic diagram \citep[e.g.,][]{kewley13} based on the [N{\sc ii}]/H$\alpha$ and [O{\sc iii}]/H$\beta$ line ratios shown in Figure~\ref{BPT}, we see that all the objects are located on the left side near the boundary separating AGN from star-forming regions. Although their location is consistent with the stellar origin, non-thermal radiation cannot be excluded. It is important to note that low-metallicity galaxies will systematically shift to the left of the BPT diagram, making it difficult to classify metal-poor AGN based on this diagram alone \citep{groves}. Finally, the broad Balmer lines in all seven objects show very steep decrements ($\gtrsim$7), which suggest either that collisional excitation is important and that the broad emission comes from very dense gas ($N_{\rm e}$$\sim$10$^{4}$ cm$^{3}$), or alternatively that the broad line region suffers from high extinction and the AGN is a type 1.9. Given that the optical continuum and emission line luminosities are consistent, the latter seems less likely.

I08 estimated BH masses of $\sim$(0.5--3)$\times$10$^{6}$\,$M_{\odot}$ for their four AGN candidates following the relation derived by GH05 between H$\alpha$ and $M_{\rm BH}$. We have similarly estimated BH masses for the three I07 objects using the GH05 relation to be $\sim$(0.4--3)$\times$10$^{5}$\,$M_{\odot}$, assuming that the broad L$_{\rm H\alpha, br}$ have an AGN origin. See Table~\ref{general} for details.

In general, the I08 AGN candidate galaxies are quite compact (marginally resolved, $\approx$2\arcsec\ in extent) and have relatively red optical colors due in part to the presence of several strong emission lines. Galaxies J1025$+$1402 and J1047$+$0739 have approximately round shapes, whereas J0045$+$1339 and J1222$+$3602 have distorted morphologies suggestive of cometary structure \citep[e.g.][]{Loose1985}. The I07 candidates are also relatively compact but have bluer optical colors likely due to their lower redshifts. Galaxies J0840$+$4707 and J1536$+$3122 have somewhat extended morphologies ($\approx$2\arcsec$\times$5\arcsec\ in extent). J1404$+$5423 is actually an individual H{\sc ii} region found in the host galaxy M101, and not an ELG. The location of this object, $\sim$11\arcmin\ ($\sim$16\,kpc) from the M101 nucleus, likely rules out a central massive black hole (hence the '?' in Table~\ref{general}). Nonetheless, we retain it throughout our analysis since its emission lines share many properties with the more luminous ELGs and may aid interpretation.

\subsection{X-ray Observations}
Based on known relations between broad H$\alpha$, [O{\sc iii}], and X-ray emission for unobscured AGN \citep[e.g.,][hereafter P06]{panessa}, we expect the I08 AGN candidates to be relatively X-ray bright emitters (e.g., with $\sim$50--300 counts in the 0.5--7.0 keV band expected in 5\,ks \textsl{Chandra} exposures). Thus, we initiated a small program to observe the four AGN candidates with the \textsl{Chandra} X-ray Observatory (PI: Thuan) using the ACIS-S CCD camera. For the three weaker AGN, we retrieved {\it Chandra} observations from the archive. The processing, screening, and analysis of the data were performed using the standard tools from CIAO (v4.4), as well as custom IDL software. No strong background flares occurred during the observations. We selected only good events (status=0) between 0.5--7.0\,keV for further study. Additional details about the observations are listed in Table~\ref{xray}. 

   \begin{table*}
   \begin{center}
   \caption[]{X-ray Properties of AGN Candidates}
       \label{xray}
       \begin{tabular}{cccccccc}
           	\hline
           	\noalign{\smallskip}
           	Object & Obs. ID & Obs. Date & Exp. & Counts & $N_{\rm H}$ & $L_{\rm 2-10 keV, obs}$ & $L_{\rm 2-10 keV, exp}$ \\
            \noalign{\smallskip}
           	\hline
           	\noalign{\smallskip}
           	J0045+1339 & 10294 & 2009 May 28 & 11562 & $<$4.6              & 5.3 & $<$5.3$\times$10$^{41}$ & 6.0$\times$10$^{42}$\\
            J1025+1402 & 10295 & 2009 Jan 14 &  4937 & $<$4.6              & 3.6 & $<$1.1$\times$10$^{41}$ & 7.3$\times$10$^{42}$\\
            J1047+0739 & 10296 & 2009 Jun 23 &  4782 & 3.0$^{+2.9}_{-1.6}$ & 2.4 &  2.2$\times$10$^{41}$ & 3.9$\times$10$^{43}$ \\
            J1222+3602 & 10297 & 2009 Feb 26 & 10718 & $<$4.6              & 1.2 & $<$5.5$\times$10$^{41}$ & 3.0$\times$10$^{42}$\\
            J1536+3122 & 11476 & 2010 Apr 05 &  1822 & 2.0$^{+2.6}_{-1.3}$ & 2.1 &  4.1$\times$10$^{40}$ & 9.1$\times$10$^{41}$ \\
            J0840+4707 & 17039 & 2015 Jan 05 & 12892 & 4.0$^{+3.2}_{-1.9}$ & 2.8 &  7.9$\times$10$^{39}$ & 1.8$\times$10$^{42}$ \\
            J1404+5423 &  2779 & 2002 Oct 31 & 14276 & 9.0$^{+4.1}_{-2.9}$ & 1.8 &  1.5$\times$10$^{37}$ & 2.7$\times$10$^{39}$ \\
\noalign{\smallskip}
           	\hline
       	\end{tabular}
     \tablefoot{{\it Col. 1:} Shortened SDSS object name. 
      {\it Col. 2:} \textsl{Chandra} observation identification number.
      {\it Col. 3:} \textsl{Chandra} observation date (UT).
      {\it Col. 4:} \textsl{Chandra} net exposure after cleaning in seconds.
      {\it Col. 5:} \textsl{Chandra} 0.5--7.0\,keV net counts derived from a 2\arcsec\ radius aperture. For detected sources, we quote Poisson errors derived from \citet{gehrels} at 1 $\sigma$ confidence level. For undetected sources, we derive 99\% confidence upper limits based on the Bayesian technique of \citet{kraft}.
      {\it Col. 6:} Galactic H{\sc I} column density from \citet{Kalberla2005} in 10$^{20}$ cm$^{-2}$.
      {\it Col. 7:} Observed 2--10 keV luminosity or limit assuming a $\Gamma$$=$1.8 powerlaw spectrum absorbed by Galactic $N_{\rm H}$, in erg s$^{-1}$.
      {\it Col. 8:} Expected 2--10 keV luminosity based on the $L_{X}$--$L_{\rm H\alpha}$ relation of P06, in erg s$^{-1}$.\label{tab3}}
   \end{center}
   \end{table*}

Among the I08 objects, only the most luminous H$\alpha$ emitter, J1047+0739, is marginally detected, with three photons at energies of 0.9, 0.95 and 2.3 keV, hinting at a soft underlying spectrum which could be ascribed to the X-ray binary population in the host galaxy. For the 27 I07 objects, a search of the archival data yields 3 detections. Given the low number of detected counts, we used the \textsl{WebPIMMS} tool to convert counts into fluxes for each particular observing cycle assuming a power-law spectrum with $\Gamma$$=$1.8 and Galactic $N_{\rm H}$, as might be expected for IMBHs. The resulting observed 2--10\,keV luminosities (or limits) are provided in Table~\ref{xray}. For count and flux upper limits, we adopt the Bayesian method of \citet{kraft}.

\subsection{Optical Spectroscopy}

Archival SDSS optical spectroscopy of our targets was previously examined in detail in I07 and I08. Subsequent observations have been acquired on a number of occasions and instruments to investigate variability as follows.

New optical spectra for galaxies J0045$+$1339, J1025$+$1402, J1047$+$0739, and J1222$+$3602 were obtained using the Apache Point Observatory (APO) 3.5m telescope using the Dual Imaging Spectrograph (DIS) in both the blue and red wavelength ranges. The sources were observed on several occasions between 2007-2015 as indicated in Table~\ref{flux}, generally under non-photometric conditions and seeing between $\sim$1--3\arcsec. Slit widths of 0.9--2\arcsec\ were used. In the blue range, we used the B400 grating, with a linear dispersion of 1.83\,\AA\,pixel$^{-1}$ and a central wavelength of 4400\AA, while in the red range we used the R300 grating with a linear dispersion of 2.31\,\AA\,pixel$^{-1}$ and a central wavelength of 7500\AA. 
Spectrophotometric standard stars Feige 34, Feige 110, and G191B2B were observed for flux calibration \citep{massey}.
The data reduction procedures are the same as described in \citet{Thuan2005}.

New optical spectra for galaxies J0045$+$1339 and J1222$+$3602 were obtained using the MODS instrument on the Large Binocular Telescope (LBT) on UT 2013 October 27 and UT 2015 May 18, respectively (Fig. ~\ref{LBT-spectra}). Slit widths of 1\arcsec\ were used. The G400L grating in the blue range yields a linear dispersion of 0.278\,\AA\,pixel$^{-1}$ and a central wavelength of 4000\AA, while the G670L grating in the red range delivers a linear dispersion of 0.173\,\AA\,pixel$^{-1}$ and a central wavelength of 8000\AA. Spectrophotometric standard stars Feige 34 and Feige 110 were observed for flux calibration \citep{massey}. For both the data reduction procedures are the same as described in \citet{Thuan2005}.

New optical spectra for galaxies J1025$+$1402 and J1047$+$0739 were obtained using the LRIS instrument on the Keck telescope on UT 2016 February 14 (Fig. \ref{keck-spectra}). The spectra were observed at the parallactic angle, through 1.5 arcsec slits, with total exposures of 600\,s each. The night was photometric, although the seeing was $\sim$1\farcs1. A 1\farcs5 wide slit was adopted, with the 5600~\AA\ dichroic to split the light, the 600 $\ell$ mm$^{-1}$ grism on the blue arm ($\lambda_{\rm blaze} = 4000$~\AA; spectral resolving power $R \equiv \lambda /\Delta \lambda \sim 1100$), and the 400 $\ell$ mm$^{-1}$ grating on the red arm ($\lambda_{\rm blaze} = 8500$~\AA; $R \sim 1000$).  The spectra were reduced and extracted following standard procedures, and flux-calibrated based on observations of spectrophotometric standard stars G191B2B and HZ44 \citep{massey}.

Line fluxes were estimated through spectral decomposition of the continuum and various emission lines. Critically, the spectra show no significant change in the broad H$\alpha$ emission lines compared to past epochs. The errors reported in Table~\ref{flux} are statistical only and do not account for additional systematic errors, which could be as large as $\sim$50\% given that many of the above observations were taken on different instruments, generally in non-photometric conditions, and with varying slit widths.

\begin{figure*}
\centering
\includegraphics[width=\hsize]{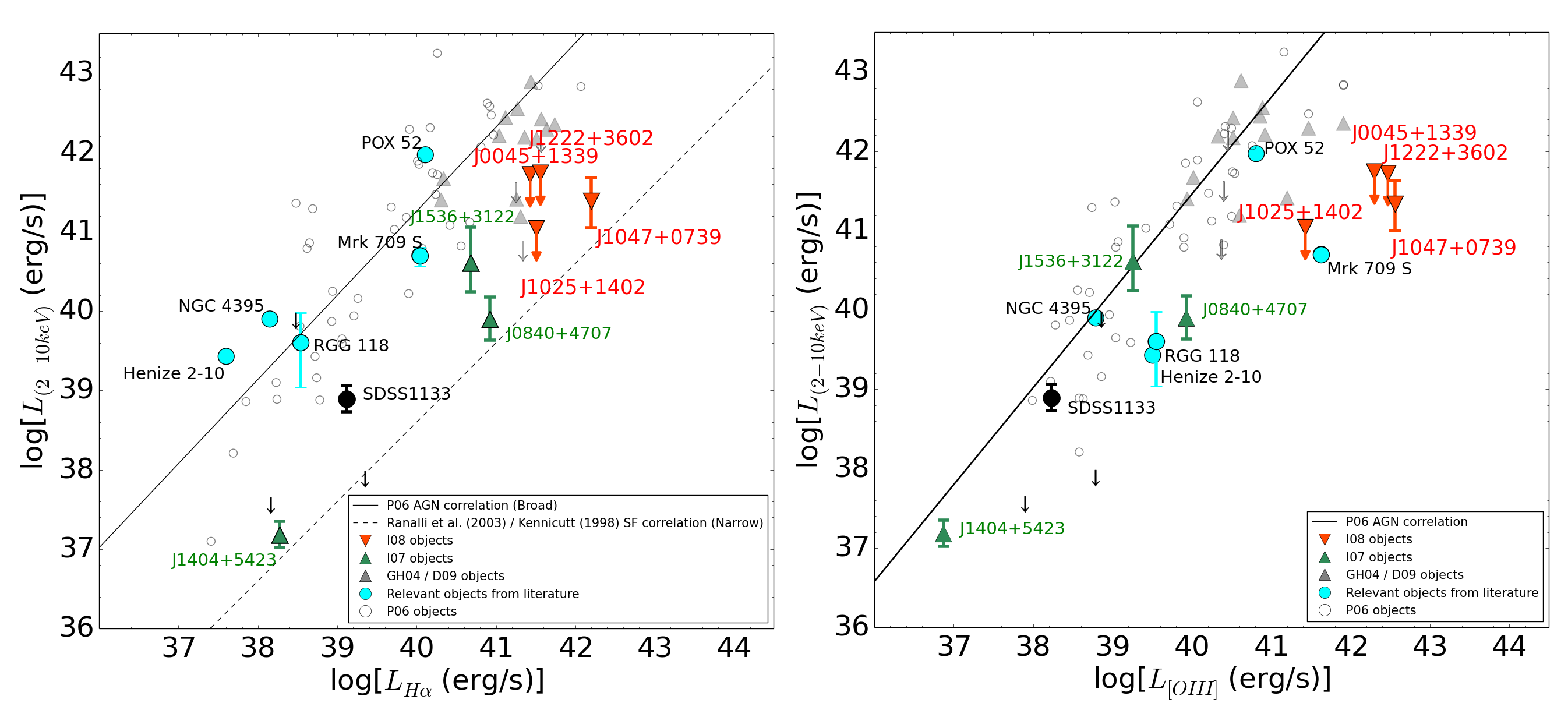}
\caption[Panessa]{Hard X-ray (2--10\,keV) emission compared to broad H$\alpha$ emission \textsl{(left)} and [O{\sc iii}] emission \textsl{(right)}. The I08 and I07 objects are denoted by red and green filled triangles, respectively. The best-fit regressions for AGN from P06 are shown as solid black lines, while the individual AGN used by P06 are shown as open black circles. The dotted line shows the hard X-ray (2--10\,keV) -- star formation rate relation presented by \cite{ranalli} and \cite{kennicutt}. Additional objects from GH04 and the literature are denoted by filled blue circles and filled grey triangles, respectively. Downward arrows denote X-ray upper limits for all samples.}
\label{Panessa-mix}
\end{figure*}

\begin{figure*}
\centering
\includegraphics[width=0.9\hsize]{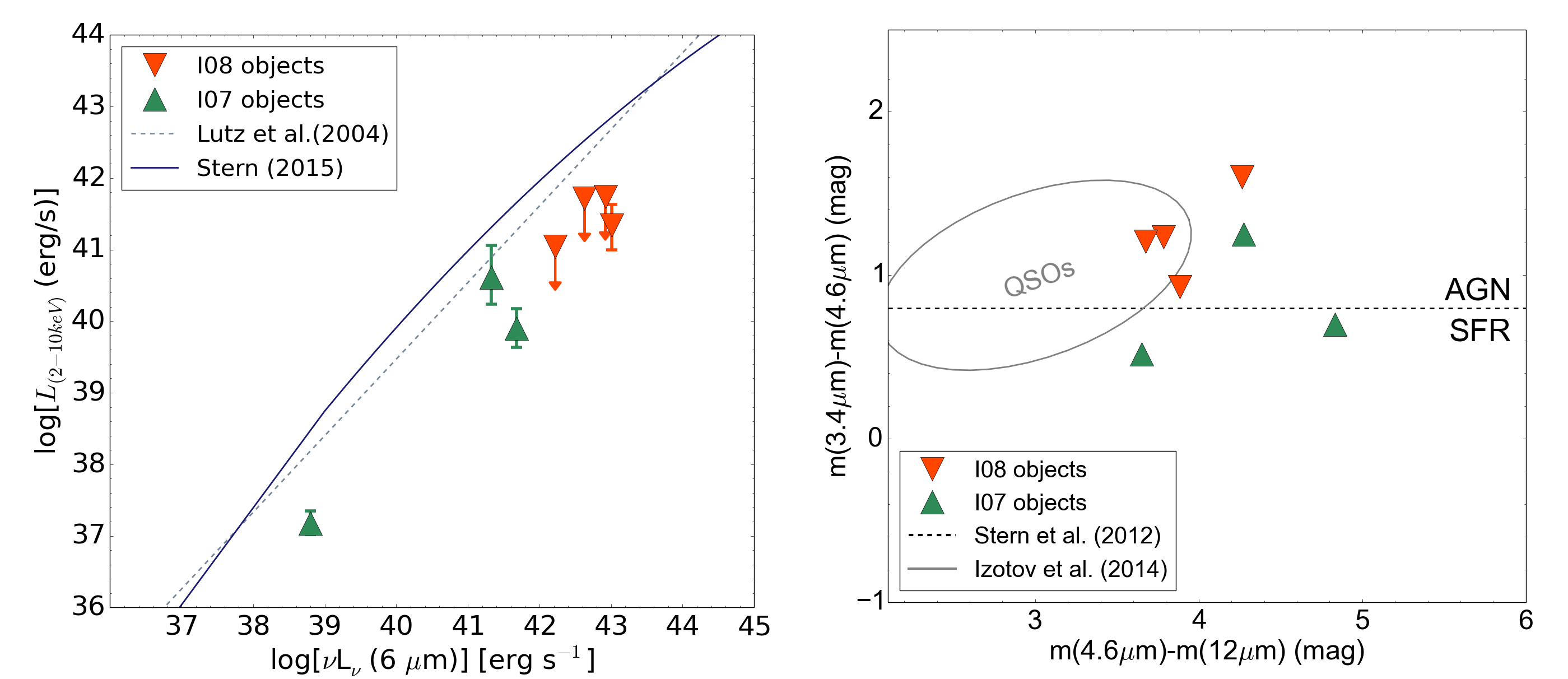}
\caption[WISE]{\textsl{(left)} Rest-frame 2--10\,keV X-ray and 6\,$\mu$m luminosities for the I08 and I07 objects, calculated following the method described in \citet{stern2015}. The dashed line indicates the best linear fit for AGN from \citet{lutz2004}, while the solid curve shows the AGN relation, corrected for high and low luminosities, from \citet{stern2015}. \textsl{(right)} Color-color diagram using the three most sensitive WISE bands 3.4\,$\mu$m ($W1$), 4.6\,$\mu$m ($W2$) and 12\,$\mu$m ($W3$). The dashed line shows the demarcation between star-forming galaxies and AGN from \citet{stern2012}. The gray ellipse represents the region where most SDSS detected QSOs lie \citep{izotov2014}. In both plots: the filled red and green triangles represent the I08 and I07 objects, respectively; downward arrows denote X-ray upper limits; and all objects are strongly detected by WISE such that their errors are smaller than the symbols representing them.
}
\label{WISE-mix}
\end{figure*}

\section{Discussion}\label{sec:discussion}
We now investigate hard X-ray (2--10 keV) constraints for the four I08 low-metallicity AGN candidates and the three \textsl{Chandra} observed galaxies from the I07 sample. Figure~\ref{Panessa-mix} compares the hard X-ray luminosity against the H$\alpha$ and [O{\sc iii}] luminosities for these objects, allowing us to examine how they lie compared to the AGN and star-forming relations of P06 and \citet{ranalli}, respectively. We also show in Figure~\ref{Panessa-mix} X-ray constraints for the P06 AGN, the GH04/D09 candidate IMBHs/SMBHs, and five additional objects: RGG 118, Pox 52, NGC 4395, Henize 2-10 and Mrk 709 (south). These last five dwarf galaxies are reported to host BHs of $\sim$10$^4$-10$^7$ \textsl{M$_{\odot}$}  \citetext{RGG 118 -- \citealp{baldassare15}; Pox 52 -- \citealp{barth04}; NGC 4395 -- \citealp{filippenko}; Henize 2-10 -- \citealp{reines12}; Mrk 709 -- \citealp{reines14}} and thus provide useful comparisons for our investigation. The BH in RGG 118 is the smallest ever reported in a galaxy nucleus, while Pox 52 and NGC 4395 are of particular interest since their BH masses ($\sim$10$^{5}$ \textsl{M$_{\odot}$}) are better constrained; all lie in the IMBH regime. Pox 52 and NGC 4395  have similar properties and are classified as dwarf Seyfert 1 galaxies. It is noteworthy that NGC 4395 has no discernible bulge, and thus is not expected to host a central BH, but its nucleus exhibits all the characteristics of Seyfert activity, including broad emission lines and X-ray variability.

The P06 AGN, GH04/D09 candidate IMBHs/SMBHs, and the five additional objects more or less all follow the expected AGN correlations, which represent various known couplings (e.g., broad and narrow line regions, X-ray corona) with the accretion disk. The only outlier appears to be the strong [O{\sc iii}] emission from Mrk 709 (south), pushing it $\sim$2 dex off the AGN relation of P06. This offset could be an X-ray deficit due to X-ray variability or obscuration, or an [O{\sc iii}] excess related to stronger AGN emission in the past. In general, the highest X-ray luminosity is shown in Fig.~\ref{Panessa-mix}, but X-ray variability still presumably contributes substantially to the dispersion in the P06 relations. Also shown are the I08 objects, denoted by downward arrows for X-ray upper limits and a red solid triangle. These categorically demonstrate a lack of X-rays for a given strength of broad H$\alpha$ and [O{\sc iii}] emission. As a sample, this is unusual and atypical of AGN. 

If the X-ray deficit is due to variability, we would expect the objects under study here to scatter around the P06 relation, which they do not. Another explanation could be strong line-of-sight obscuration ($N_{\rm H}$$\gtrsim$$10^{24}$ cm$^{-2}$), with the caveat that the few photons that are detected in J1047+0739 are low-energy ones. However, assuming this geometry for all four I08 objects and a standard AGN orientation paradigm might imply a large population of unobscured AGN in dwarf galaxies that is not currently observed. A third possibility is that these objects are intrinsically X-ray weak, whereby the characteristic strong X-ray emitting corona is never produced. This odd behavior has been observed in a few Broad Absorption Line (BAL) quasars and luminous infrared galaxies \citep[e.g.,][]{Luo2014, Teng2015}, although the I08 objects ought to occupy a very different physical regime. Given the lack of understanding in these objects, this possibility remains difficult to exclude, however.

    We further investigate the nature of the X-ray deficit by examining the mid-infrared (MIR) emission for the sample, benefiting from the Wide-field Infrared Survey Explorer (WISE), which imaged the entire sky in four bands centered at 3.4, 4.6, 12 and 22 $\mu$m, called $W1$, $W2$, $W3$ and $W4$, respectively. If these I08 and I07 objects are highly obscured AGN (e.g., a large fraction of the accretion disk emission is blocked by a dusty torus), we may see their re-radiated emission at MIR wavelengths \citep[e.g.,][]{efstathiou1995}, since this hot dust emission is largely unaffected by further obscuration from the torus or interstellar medium. AGN are often revealed either by their characteristic X-ray-to-MIR ratios \citep{lutz2004, gandhi2009, asmus2015, stern2015} or red MIR colors \citep[e.g.,][]{richards2006, assef2010}. The X-ray-to-MIR ratio is well suited for detecting obscured AGN, since the X-ray emission is expected to be suppressed compared to the MIR emission \citep[e.g.,][]{alexander2008, lanzuisi2009, bauer2010}. Likewise,  \citet{stern2012} found that a cut at $W1$$-$$W2$$\geq$0.8\,mag provides a very reliable AGN indicator of hot dust in both obscured and unobscured sources with $W2$$<$15.05\,mag, based on a WISE-selected sample of AGN.

Figure~\ref{WISE-mix} shows the mentioned relations for the I08 and I07 objects from Table~\ref{tab3}, represented by red and green triangles, respectively, with upper limits denoted by downward arrows. On the left side, we compare the rest-frame hard 2--10 keV X-ray and 6\,$\mu$m luminosities of the objects against the standard relations from \citet{lutz2004} and \citet{stern2015}. Notably, all the galaxies lie below the AGN relations, most by factors of at least $\sim$10--60 (three sources have only X-ray upper limits and could be substantially lower still). This is consistent with our previous findings of X-ray deficits. On the right side, we show the observed MIR colors $W1$$-$$W2$ and $W2$$-$$W3$ for the I08 and I07 objects, where the dashed line denotes the $W1$$-$$W2$$\geq$0.8\,mag limit from \cite{stern2012}. Five objects lie above the line, suggesting an AGN classification. Critically, however, \citet{izotov2014} examined the MIR colors of $\sim$10,000 star-forming galaxies with strong emission lines and no obvious signs of AGN, detected in both SDSS and WISE, and found that a non-negligible number ($\sim$5\%) scattered above the fiducial AGN criterion of \citet{stern2012}, although they remained distinctly offset in $W2$$-$$W3$ from QSOs \citep[e.g., Figure~7c of][]{izotov2014}. \citet{izotov2014} found that these objects are mainly luminous galaxies with high-excitation H{\sc ii} regions, and their unusual WISE colors are produced by hot dust associated with radiation from young star-forming regions. Meaning that the I08 and I07 objects could belong to this tail of the star-forming galaxy population. Thus, some objects exhibit MIR properties consistent with highly obscured AGN activity, but we lack conclusive results that can differentiate AGN from star-forming regions for our samples.

Notably, most of the I08 objects have strong limits on high-ionization emission lines like He{\sc ii} and [Ne {\sc v}], implying that there is also no strong non-thermal hard ionizing UV radiation present. The exception is J1222$+$3602, which has a [Ne {\sc v}]/He{\sc ii} ratio of $\sim$1.5. This is $\sim$5 times higher than in star-forming galaxies and is more consistent with the value for Seyfert 2 galaxies. In general, this is consistent with the observed deficit of X-ray emission, and implies that the BHs in the I08 objects may not be strongly accreting at all, or as I08 argued, that there is a high covering factor from the accretion disk itself that absorbs the hard UV (and now X-ray) emission. One final interesting scenario is for the circumnuclear gas to be influenced by the gravitational potential of a dormant IMBH/SMBH but excited by stellar processes. Strong narrow emission lines are present in all I08 objects, implying that stellar processes are able to generate sufficient photon or shock excitation. It could be possible that these processes are energetic enough to power the broad line excitation too. In this scenario, most of the ionizing stars or shocks would presumably need to reside close to the broad line gas in order to satisfy the observed UV continuum constraints. This seems like a physically implausible scenario. 

\begin{figure*}
   \centering
   \includegraphics[width=16cm]{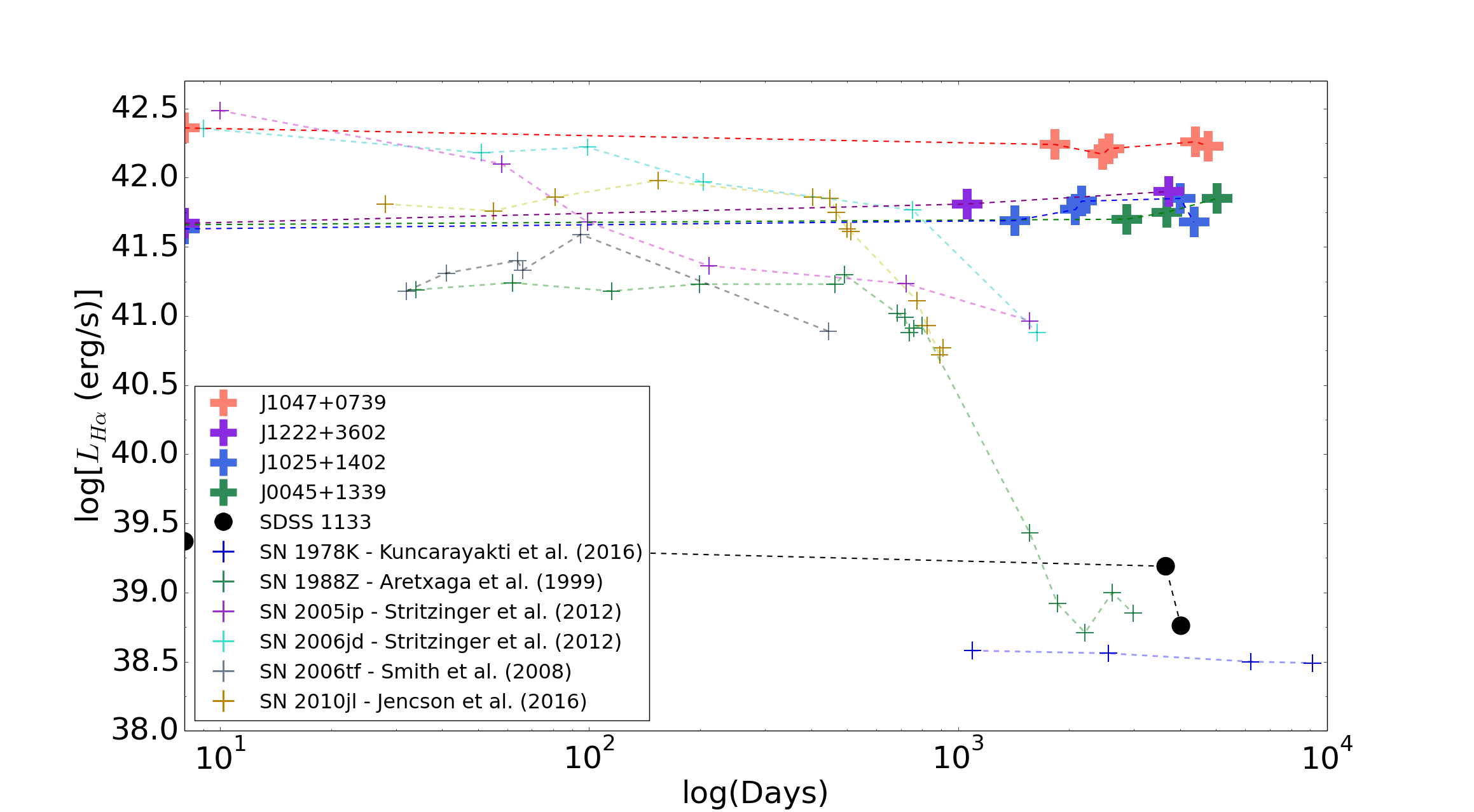}
      \caption[SNe]{Time evolution of the broad H$\alpha$ luminosity for the I08 objects, as well as several luminous type IIn SNe \citep{kuncarayakti, aretxaga, stritzinger, smith, jencson} and the transient event SDSS1133 \citep{koss}. The I08 objects are roughly constant over periods of 10--13 yr; there is some variability between epochs at the $\sim$50\% level, but this could be due to calibration uncertainties. All error bars are smaller than the symbols.}
     \label{SNe}
   \end{figure*}

One of the mechanisms proposed by I08 to explain the observed offset in broad H$\alpha$ is the presence of type IIn SNe, since this type of SNe can produce relatively long-lived, high broad H$\alpha$ luminosities. Figure~\ref{SNe} shows the time evolution for several luminous type IIn SNe compared to the I08 objects. The I08 objects show no significant variation in broad H$\alpha$ emission at least over periods of $\sim$10--13 years. While a few individual SNe have been observed to have broad H$\alpha$ luminosities as high as the I08 objects \citep[e.g., SN 2005ip and SN 2006jd;][]{stritzinger}, Figure~\ref{SNe} shows that such powerful SNe can only generate such luminosities for $\lesssim$3\,yr. SN 1978K stands out for having a broad H$\alpha$ luminosity that has shown little variation over the course of $\sim$25 years; however, its luminosity is 2--3 dex less than those of the I08 objects. Thus, the I08 objects would require large numbers of such SNe to maintain the bright, roughly constant broad H$\alpha$ luminosities that are observed. Because such SNe are thought to be produced by high mass stars, this scenario seems unlikely based on the star formation rates, initial mass function, and general lack of transients observed in such dwarfs.

\citet{koss} reported on an unusually persistent transient, SDSS J113323.97$+$550415.8 (hereafter SDSS1133), in the nearby blue compact dwarf galaxy Mrk 177, which shares some similarities with the I08 objects. Unlike the I08 objects, the transient lies 5\farcs8 away from the apparent nucleus (a projected offset of 0.8\,kpc) and has slowly varying broad Balmer line emission with a velocity offset of $\sim$200--800 km\,s$^{-1}$ from the host nucleus and a peak luminosity of $\sim 10^{40}$ erg\,s$^{-1}$. Such properties imply that SDSS1133 could be a black hole recoil candidate, although some traits can also be explained by a luminous blue variable star that was erupting for decades before exploding as a type IIn SN. With respect to the latter, Koss et al. argue that this event would represent one of the most extreme episodes of pre-SN mass-loss ever discovered. Mrk 177 shows signs of clumpy star formation, analogous to the extended emission observed for two I08 galaxies. The emission lines from the Mrk 177 host nucleus ($Z$$\approx$$Z_{\odot}$) and SDSS1133 generally place both in the star-forming galaxy region of the diagnostics plots, although at its centroid SDSS1133 shifts somewhat into the Seyfert galaxy regime. SDSS1133 also shows 1--3 magnitudes of optical variability on timescales of $\sim$1--60 years, unlike the I08 objects. It is interesting to consider that if SDSS1133 were coincident with the nucleus, it would be more strongly contaminated by the nuclear star formation and vary less, and therefore be even more comparable to the I08 objects. Intriguingly, SDSS1133 is marginally detected by \textsl{Swift} XRT with 7.6 $\pm$ 3.4 background-subtracted 0.3--10 keV counts, corresponding to a 0.3–-10 keV luminosity of 1.5$\times$10$^{39}$ erg s$^{-1}$ (adopting a power-law spectrum with fixed $\Gamma=1.9$). With this limited amount of X-ray emission, Koss et al. cannot distinguish between the AGN and SN scenarios, similar to the case of J1047+0739. The broad H$\alpha$ (from 2003) and X-ray (from 2013) luminosities of SDSS1133 place it $\sim$1.5 dex off the P06 AGN relation, similar to the I08 objects, although these were not taken contemporaneously and the X-ray emission could have been significantly higher 10 years beforehand.
  
\section{Summary}

We have studied the X-ray, H$\alpha$, and [O {\sc iii}] emission properties from a sample of low-metallicity compact dwarf galaxies discovered by SDSS to have unusually broad and luminous H$\alpha$ emission.
   \begin{enumerate}
      \item Given the strength of the broad H$\alpha$ and total [O {\sc iii}] emission, if they were produced by an accretion disk surrounding an IMBH/SMBH, we would expect to detect significant X-ray emission for all the I07 and I08 objects. However, surprisingly, only one of the four I08 objects - J1047$+$0739 - has a (marginal) X-ray detection, where we find three soft (0.9--2.3 keV) photons in 4782 seconds. The latter can be explained as soft emission produced by the X-ray binary population in the host galaxy. Regarding the I07 objects, we found 3 Chandra detections in the archives. For these, a pure stellar origin of the broad H$\alpha$ luminosities cannot be ruled out. 
      \item Our X-ray limits constrain the I07 and I08 objects to lie at least $\sim$1--2 dex below the known AGN relations. These limits are consistent with the apparent lack of any strong non-thermal hard ionizing UV radiation suggested by the weak high-ionization emission lines. Together, the X-ray and UV constraints imply that no strong accretion onto IMBH/SMBHs is present in these objects, or that alternatively it is completely obscured along our line of sight.
      \item Continued spectral monitoring of the broad H$\alpha$ emission lines in the I08 objects demonstrates that their broad lines have remained approximately constant for at least $\sim$10--13 years. The production of such long-lived, high luminosity broad lines is strongly incompatible with individual known SNe, and would require many 10s to 100s of such SNe in order to sustain their luminosity and lack of strong variability. This scenario seems unlikely based on the host galaxy properties and low number of transients observed in similar compact dwarfs to date.
   \end{enumerate}

While the I07 objects conform in many ways to the characteristics of star-forming galaxies, the I08 objects have a number of intriguing properties which are non-trivial to explain. If these objects are in fact AGN as proposed originally, they imply that X-ray and UV weak AGN exist. It is unclear how the AGN relations might behave in the low BH mass, low-metallicity regime, and confirmation of BHs in these objects could thus provide possible insights on AGN activity in metal-poor dwarf galaxies. On the other hand, if these objects are not AGN, then their nature is uncertain and perplexing to explain in the context of current empirical and theoretical properties of star formation and SNe, and would likely require extreme events or processes that are as yet unknown. More generally, if such broad lines can be produced via stellar processes, the I08 objects highlight a clear flaw in identifying and characterizing lower mass AGN based on their broad optical emission lines. This would leave many of the BH mass estimates made over the past decade on shaky ground. In this sense, the I08 objects remain highly intriguing laboratories for future study. 

The favored explanation - AGN - cannot be firmly confirmed or discarded at this point, and more data are necessary to understand the underlying phenomenon in these galaxies. In particular, focused studies of the most extreme I08 objects at X-ray, optical, and radio wavelengths will help to understand whether they can still be accommodated under the AGN paradigm; VLA and EVN observations are currently being analyzed (N. Guseva et al. in preparation). Longer X-ray exposures of J1047+0739 will characterize better its X-ray spectral and temporal nature. High spatial resolution imaging and spectroscopy can place better constraints on the spatial extent of the broad emission lines and determine what fraction might be attributable to stellar processes. And finally, Very Long Baseline Array observations can help pinpoint the location and properties of any non-thermal sources buried in the centers of these compact galaxies. These studies would help to address how robust optical broad lines are as tracers of AGN activity, a robustness that has been called into question by our current constraints on the I08 objects.

\begin{acknowledgements}
We thank Mislav Balokovic for help acquiring Keck spectra. The work of Daniel Stern was carried out at the Jet Propulsion Laboratory, California Institute of Technology, under a contract with NASA.
We acknowledge support from 
CONICYT-Chile grants Basal-CATA PFB-06/2007 (FEB), FONDECYT Regular 1141218 (CS, FEB), and "EMBIGGEN" Anillo ACT1101 (FEB);
the Ministry of Economy, Development, and Tourism's Millennium Science
Initiative through grant IC120009, awarded to The Millennium Institute
of Astrophysics, MAS (FEB); and 
NASA through Chandra Award Number GO9-0106C (FEB, TXT) issued by the Chandra X-ray Observatory Center, which is operated by the Smithsonian Astrophysical Observatory for and on behalf of the NASA under contract NAS8-03060.
\end{acknowledgements}


\pagebreak
\begin{appendix}

\section{Optical Spectra}\label{appendix}

\begin{figure*}
\centering
\includegraphics[width=\hsize]{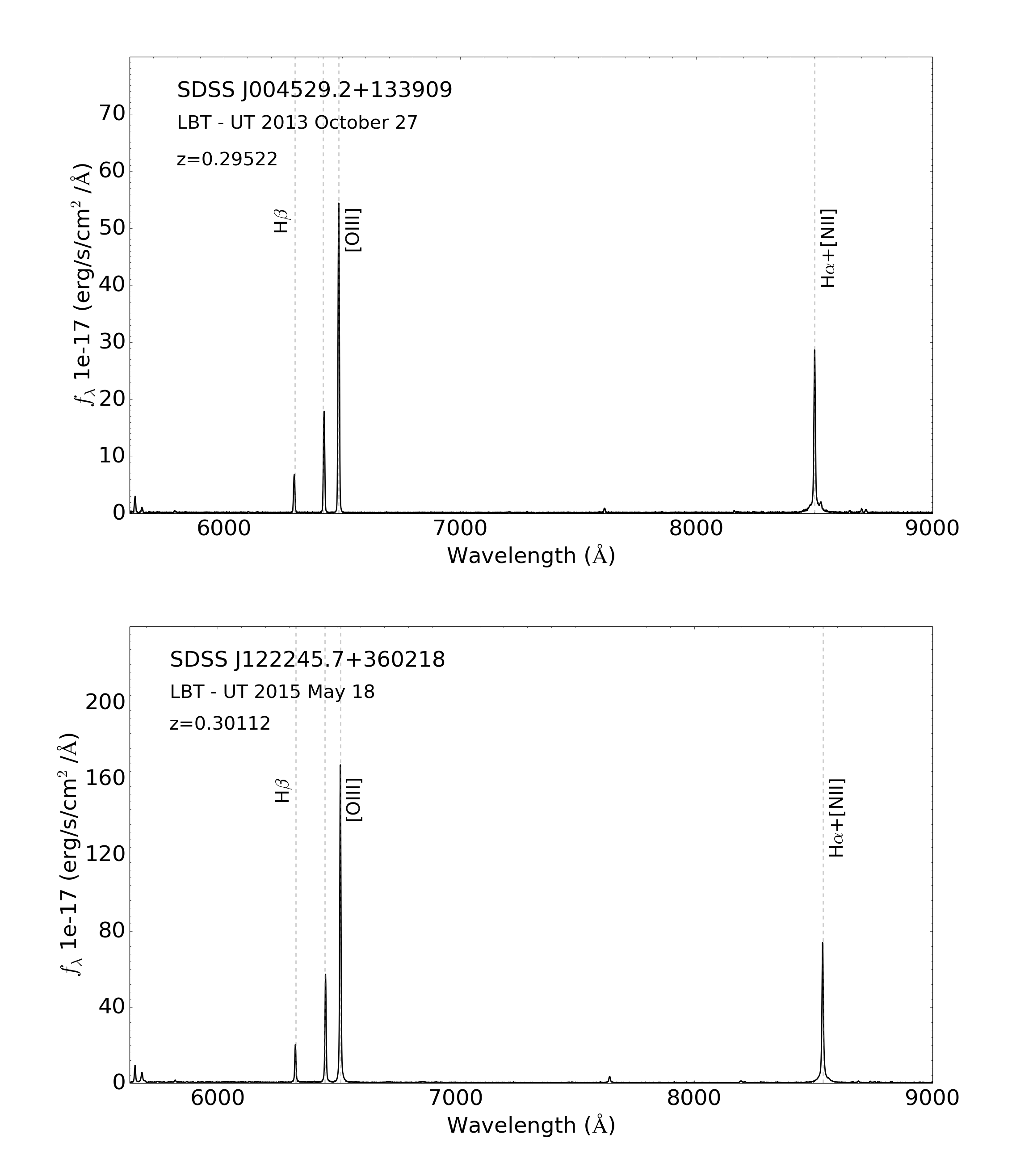}
\caption[LBT]{Optical spectra for galaxies J0045+1339 and J1222+3602 obtained using the LBT telescope on UT 2013 October 27 and UT 2015 May 18, respectively. The broad H$\alpha$ flux does not present considerable variations compared to previous measurements.}
\label{LBT-spectra}
\end{figure*}

\begin{figure*}
\centering
\includegraphics[width=\hsize]{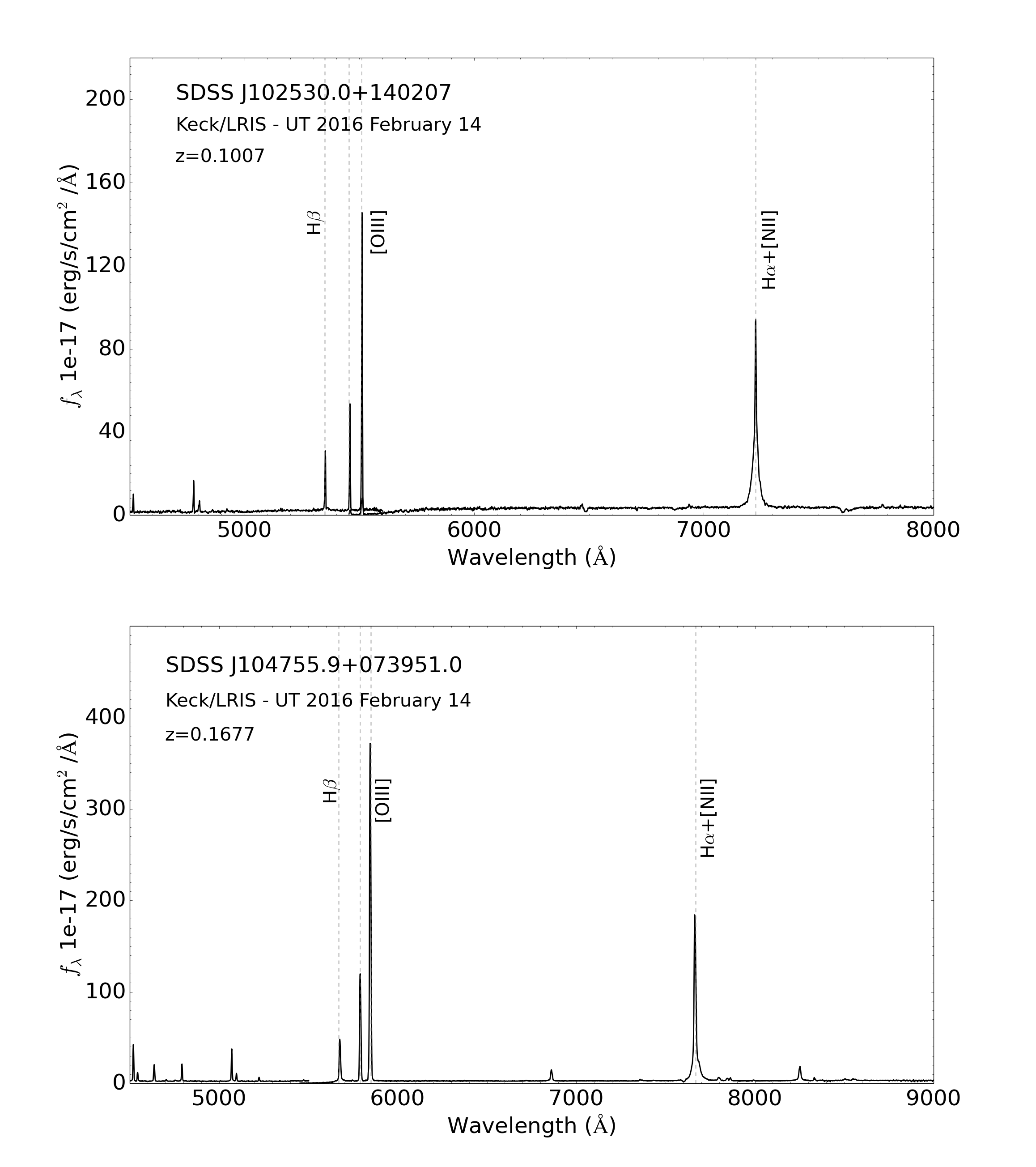}
\caption[Keck]{Optical spectra for galaxies J1025+1402 and J1047+0739 obtained using the LRIS instrument on the Keck telescope on UT 2016 February 14. The broad H$\alpha$ flux does not present considerable variations compared to previous measurements.}
\label{keck-spectra}
\end{figure*}

\end{appendix}

\end{document}